# NEOPROP: A NEO PROPAGATOR FOR SPACE SITUATIONAL AWARENESS FOR THE 6TH IAASS CONFERENCE

**Valentino Zuccarelli [1], David Bancelin[2], Sven Weikert[3], William Thuillot[4], Daniel Hestroffer[5], Celia Yabar Valles [6], Detlef Koschny[7]**

[1] Astos Solutions GmbH, Grund 1, 78089 Unterkirnach (Germany), Email: Valentino.Zuccarelli@astos.de
[2] IMCCE-Observatoire de Paris, Av. Denfert-Rochereau 77, 75014 Paris (France), Email: David.Bancelin@imcce.fr
[3] Astos Solutions GmbH, Grund 1, 78089 Unterkirnach (Germany), Email: Sven.Weikert@astos.de
[4] IMCCE-Observatoire de Paris, Av. Denfert-Rochereau 77, 75014 Paris (France), Email: William.Thuillot@imcce.fr
[5] IMCCE-Observatoire de Paris, Av. Denfert-Rochereau 77, 75014 Paris (France), Email: Daniel.Hestroffer@imcce.fr
[6] ESA-ESTEC, Keplerlaan 1, 2201 Noordwijk, (The Netherlands), Email: Celia.Yabar.Valles@esa.int
[7] ESA-ESTEC, Keplerlaan 1, 2201 Noordwijk, (The Netherlands), Email: Detlef.Koschny@esa.int

**ABSTRACT**

The overall aim of the Space Situational Awareness (SSA) Preparatory Programme is to support the European independent utilisation of and access to space for research or services, through providing timely and quality data, information, services and knowledge regarding the environment, the threats and the sustainable exploitation of the outer space surrounding our planet Earth. The SSA system will comprise three main segments:
- Space Weather (SWE) monitoring and forecast
- Near-Earth Objects (NEO) survey and follow-up
- Space Surveillance and Tracking (SST) of man-made space objects

Currently, there are over 600.000 asteroids known in our Solar System, where more than 9.500 of these are NEOs. These could potentially hit our planet and depending on their size could produce considerable damage. For this reason NEOs deserve active detection and tracking efforts.

The role of the SSA programme is to provide warning services against potential asteroid impact hazards, including discovery, identification, orbit prediction and civil alert capabilities. ESA is now working to develop a NEO Coordination Centre which will later evolve into a SSA-NEO Small Bodies Data Centre (SBDC), located at ESA/ESRIN, Italy. The Software prototype developed in the frame of this activity may be later implemented as a part of the SSA-NEO programme simulators aimed at assessing the trajectory of asteroids.

There already exist different algorithms to predict orbits for NEOs. The objective of this activity is to come up with a different trajectory prediction algorithm, which allows an independent validation of the current algorithms within the SSA-NEO segment (e.g. NEODyS, JPL Sentry System).

The key objective of this activity was to design, develop, test, verify, and validate trajectory prediction algorithm of NEOs in order to be able to compute analytically and numerically the minimum orbital intersection distances (MOIDs).

The NEOPROP software consists of two separate modules/tools:
1. The Analytical Module makes use of analytical algorithms in order to rapidly assess the impact risk of a NEO. It is responsible for the preliminary analysis. Orbit Determination algorithms, as the Gauss and the Linear Least Squares (LLS) methods, will determine the initial state (from MPC observations), along with its uncertainty, and the MOID of the NEO (analytically).
2. The Numerical Module makes use of numerical algorithms in order to refine and to better assess the impact probabilities. The initial state provided by the orbit determination process will be used to numerically propagate the trajectory. The numerical propagation can be run in two modes: one faster ("*fast analysis*"), in order to get a fast evaluation of the trajectory and one more precise ("*complete analysis*") taking into consideration more detailed perturbation models. Moreover, a configurable number of Virtual Asteroids (VAs) will be numerically propagated in order to determine the Earth closest approach. This new "MOID" computation differs from the analytical one since it takes into consideration the full dynamics of the problem.

## 1. ARCHITECTURE

The general architecture of the tool, with special focus on the external and internal interfaces between the different components, is assessed here.

The high-level input/output interaction of the Analytical Module can be summarized as follows:
- As input, this module reads from an external file a list of observations;
- As output, it produces a file with the determined initial state and uncertainties of the NEO and the computed MOID;

- As optional output, this module can produce a file containing a list of VAs, computed on the Line Of Variation (LOV).

The high-level input/output interaction of the Numerical Module can be summarized as follows:

- As input, this module reads the main settings from an XML file (e,g, physical properties of the NEO, final epoch, integrator, etc). Moreover, it uses the output file created by the Analytical Module (containing the determined initial state) in order to initialize the propagation;
- As optional input, it can import a list of VAs initial states, precomputed by the Analytical module;
- As output, this module produces a file with the propagated nominal trajectory and a separate output file with the list of close approaches to any planet. If VAs trajectories are propagated, then an additional output file provides their trajectory data;
- As optional output, it can produce a file with the relative positions of the NEO w.r.t. the third-bodies used during the propagation. This file can be particularly useful if the trajectory has to be visualized.

The described interfaces are summarized in Figure 1. NEOPROP Interfaces (note that the dashed lines are used for optional functionalities/files).

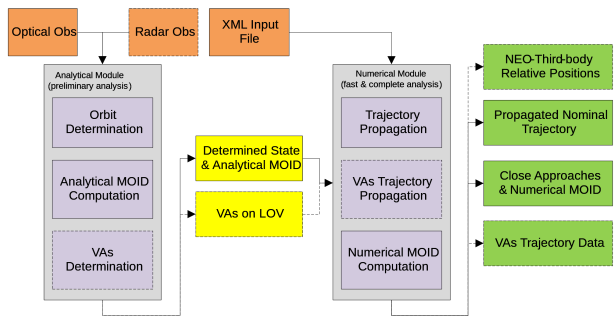

*Figure 1. NEOPROP Interfaces*

Having two stand-alone modules separates the two main tasks: orbit determination and trajectory propagation, allowing great flexibility. In this way, the different functionalities can be combined in order to have different analysis. Moreover, the integration of this tool within the NEO SSA Pilot Data Centre will be straightforward (also thanks to the lack of dependencies from external tools). This approach allows also to call the Analytical and the Numerical Modules at the same time for two different objects.

Therefore, in order to perform the risk assessment of the full NEOs catalogue (e.g. in the SBDC), only the Analytical Module will have to be run for all the objects. A list of PHAs will be determined, depending on the value of the computed MOID (and absolute magnitude). If the value of the MOID is critical (e.g. below 0.05 AU), the initial state will be numerically propagated, along with VAs, in order to evaluate the dynamical effects on the MOID.

When new observations will be available for a certain object, only the analytical module will have to be run again.

In the next figure the high-level NEOPROP architecture is shown:

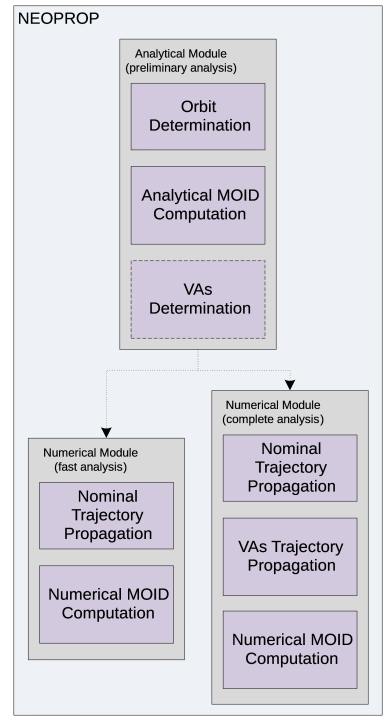

*Figure 2. NEOPROP Architecture*

A general aspect that should be highlighted is that both the Analytical and the Numerical modules support multi-threading. The tool is able to detect how many threads are available and to assign them in order to parallelize some computations. In the next paragraphs more details are provided.

## 2. ANALYTICAL MODULE

The Analytical Module allows determining the initial state (with uncertainty) of a NEO from a set of observations.

First a Gauss algorithm is run in order to get a preliminary solution of the orbit determination problem. For this solution only optical observations are used. If the test case has been run already once and a solution already exists, then the preliminary solution is not recomputed, but it is imported (along with the already used observations) and improved with the LLS technique by means of the new set of observations.

The Gauss method needs only three observations in order to compute a preliminary initial state. The first, last and intermediate observations are by default the

first ones to be used. Then the tool runs the following combinations:
- The *first* and the *last* observations are kept fixed, while the third one is moved along the observational arc;
- "forward loop": the *last* observation is used with any other two;
- "backward loop": the *first* observation is used with any other two.

Before accepting any solution, the tool checks if the determined orbital elements are compatible with the orbit of a NEO (perihelion < 1.3 AU). The algorithm stops when a minimum RMS (predefined by the user) is reached or when all the combinations have been explored. In order to speed up this time-consuming process, a multi-threading approach has been implemented. In this way several combinations (depending on the available cores) are checked at the same time.

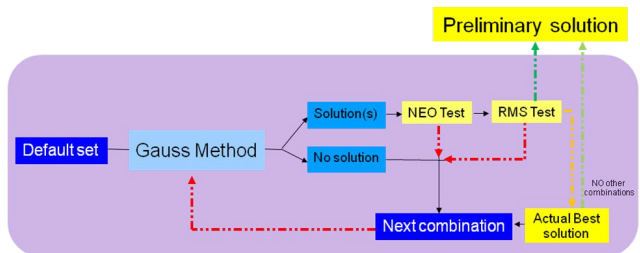

*Figure 3. Gauss method algorithm*

Then the preliminary solution is improved by means of a LLS technique, which eventually considers also radar observations. If requested, the initial states of a set of VAs are computed along the LOV (two threads can be used for this computation). Finally, the determined initial state of the NEO is used in order to analytically compute its MOID by using the Sitarski method.

The described work logic is showed by the following diagram:

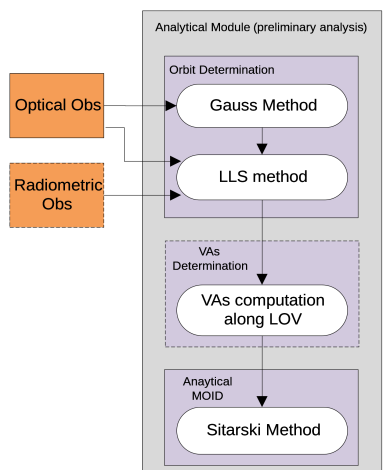

*Figure 4. Analytical Module detailed architecture*

## 3. NUMERICAL MODULE

The Numerical module performs numerical integrations. Its "fast analysis" is supposed to preliminary assess the trajectory of a NEO. When the MOID computed by the Analytical module is small enough to be potentially dangerous for the Earth, numerical integration is recommended. During this fast analysis, the initial state provided is numerically propagated and the close approaches computed.

Even though the "fast analysis" shares the same models and XML input file with those used in the "complete analysis", it does not allow the same flexibility. For instance, a fixed set of perturbations is used during this analysis, while in the "complete analysis" the user can freely customize them. This approach has been chosen for two reasons:
1. The fast analysis is supposed to be "fast". For this reason only some basic (and most important) perturbations are considered;
2. The fast analysis may be run even by users not so familiar with numerical propagations. For this reason many settings are initialized by default.

The "complete analysis" is intended to be used for a complete risk assessment of a NEO. Starting from the orbit determined by the Analytical module and from a set of VAs, the nominal trajectory will be propagated and the "dynamical" MOID computed. VAs trajectories will be also propagated in order to give to the user some more information about the sensitivity of the MOID due to uncertainties in the initial conditions. The initial states of the VAs can be computed by the Analytical Module along the LOV or by a Monte Carlo run. In this way a mix of random VAs and VAs along the LOV can be propagated.

In the Numerical module, the propagator reads the input from XML file and initializes all the models (perturbations, integrator, asteroid, etc). Then the numerical integration is started using as input the initial state of an external file provided by the Analytical module. During the propagation, all close approaches with any planets are computed and stored. If the Numerical module is run in the "complete analysis" mode, then the VAs are initialized to be ready for the propagation. Then, the propagator is called again for each VA that needs to be propagated. If a multi-core processor is used, VAs can be propagated in parallel, since these computations are independent of each other.

The work logic of the Numerical module is shown in the next figure:

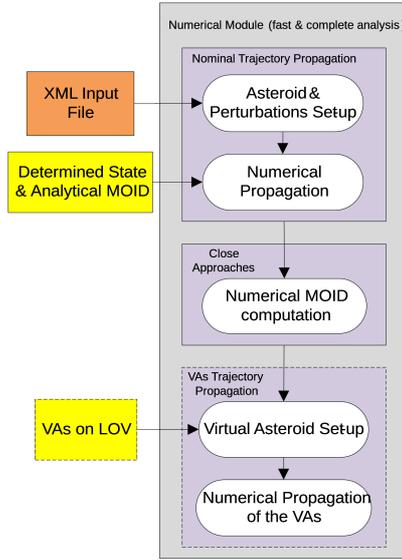

*Figure 5. Numerical Module detailed architecture*

The following external libraries/database are used in this module: Spice ephemeris, asteroid orbits database (Astorb), time standards conversion tables (IERS), Earth orientation data (IERS), observatories database including their accuracy (MPC) and Earth spherical harmonics (EGM96).
In Table 1 and Table 2, the implemented integrators and perturbation models are listed.

*Table 1. NEOPROP implemented integrators*

| Integrator | Single/Multi-Step | Step-Size |
|---|---|---|
| **Runge-Kutta 45** | single | variable |
| **Dormand Prince 8** | single | variable |
| **Runge-Kutta 853** | single | variable |
| **Runge-Kutta 4** | single | fixed |
| **Runge-Kutta 4** *Adapted* | single | fixed* |
| **Gauss-Jackson 8** | multi | fixed |
| **Gauss-Jackson 8** *Adapted* | multi | fixed* |
| **Gauss-Jackson 8** *Self-Adapted* | multi | fixed* |

*The integration follows a fixed step-size scheme, but for some trajectory arcs (e.g. close to a celestial body) the step-size might be reduced by a factor of 10.

The term "adapted" has been used to distinguish the original integrator scheme from a new algorithm implemented in NEOPROP to improve the performance. The "adapted" integrators are fixed step-size integrators which are able to reduce the step-size whenever a certain condition applies. For the "adapted" integrators, the step-size is reduced by a factor of 10 when the NEO exits the sphere of influence of the Sun. This algorithm has been implemented, because in these arcs the perturbing accelerations become significant and in order to improve the accuracy a smaller step-size is required. On the other hand, along most of the trajectory (when gravity perturbations are not so relevant) larger step-size gives better performance. The Gauss-Jackson 8 "self-adapted" integrator reduces the step-size when the ratio between the acceleration due to perturbations and the total acceleration is larger than 0.0001.

Many physical properties can be specified to characterize the NEO: its mass, absolute magnitude, diameter, bulk and surface density, thermal conductivity, thermal capacity, albedo, infrared emissivity, spin axis obliquity and rotational period (no tumbling or binary asteroid so far considered). At least the mass and the diameter or the absolute magnitude must be provided (the others can be set by default). These parameters affect the solar radiation pressure perturbation, while the others affect also the Yarkovsky computation.
The list of implemented perturbations for the "fast" and "complete" analysis is summarized in Table 2.

*Table 2. Implemented perturbation models*

| "Fast" Analysis | "Complete" Analysis |
|---|---|
| Third-Body (planets) | Third-Body (planets + 4 main asteroids) |
| - | Relativistic Effects |
| Solar Radiation Pressure | Solar Radiation Pressure |
| - | Yarkovsky |
| Earth Spherical Harmonics | Earth Spherical Harmonics |

### 4. TEST CASES

#### 4.1. Analytical Module

The Analytical Module has been tested with 6 different asteroids: the orbital elements and the MOID have been computed at certain reference epochs. As input the available observations (optical and radar) before the reference epoch has been used. As reference the NEODyS [2] system was used.
The 6 NEOs chosen for this test case are: 2011AG5, 2012DA14, 2007VK184, 2009FJ, 2004DC and (99942) Apophis. These have been selected because they all have a small MOID (below 0.01 AU) and because they cover different types of orbits. Moreover, for the NEOs 99942, 2004DC and 2012DA14 also radar observations are available.
For the asteroid (99942) Apophis 1681

optical observations and 7 radar measurements have been used.

In order to have comparable results, in NEOPROP the output has been explicitly requested at the reference epoch 56400.0 MJD (= 2456400.5 JD). This causes a small loss in accuracy, since the initial state is determined at a certain epoch (middle observation used in the Gauss method) and then it is propagated up to the requested epoch.

The orbital fitting has been executed in two different runs. This approach is also followed by the Orbfit tool, which is used in NEODyS. This type of approach is required when many observations are available and when they cover a very long observational arc. In this test case we have 1688 observations spread over a period of almost 10 years. The "multi-step fitting" strategy allows to initially constrain the orbit using a smaller set of observations and then to further refine and improve the determined initial state by using the full set of observations.

For the NEOs 2011AG5, 2007VK184 and 2009FJ, all the available observations (222, 102 and 131) are used at once. The preliminary solution and orbital fit are performed in a unique run.

2004DC requires a 2 steps orbital fit (164 optical and 8 radar observations for the first run and other 35 optical for the second run). This was required due to the fact that these 207 observations cover a very long arc (almost 8 years).

2012DA14 requires a 3 steps orbital fit (191 optical for the first run, 153 for the second one and other 572 optical and 7 radar observations for the last run). This approach is due to the close approach that this NEO had with the Earth on the 15$^{th}$ of February 2013 [1]. This close approach changed consistently its orbital elements. Moreover, a lot of new observations were taken during it. For this reason the orbital fit process became quite complex. In this test case a reference epoch before this close approach date has been chosen. In this way the MOID computation should be able to assess the close approach distance reached on the 15$^{th}$ of February.

These last two NEOs together with Apophis allow simulating a re-computation of the orbital parameters due to new available observations. This is the typical scenario the tool will have to deal with. Only when a NEO is newly discovered, the total set of observations are used together (single or multiple runs) to generate a first solution. Afterwards every time new observations are taken, the solution is just updated by using the new measurements.

In the next two tables, the results obtained by NEOPROP are compared to the reference data.

*Table 3. NEODyS-NEOPROP results comparison (99942, 2009FJ, 2011AG5)*

|  |  |  |  |
|---|---|---|---|
| *Epoch [JD]* | 2456400.5 | 2456200.5 | 2456400.5 |
| Δ-a [AU] | 0.0000000 | 0.0003387 | 0.0000001 |
| %-a | 0.000% | 0.015% | 0.000% |
| Δ-e [-] | 0.0000000 | 0.0000676 | 0.0000001 |
| %-e | 0.000% | 0.012% | 0.000% |
| Δ-i [°] |  | -0.0016318 | 0.0012234 |
| %-i | 0.047% | -0.184% | 0.033% |
| Δ-Ω [°] |  | -0.0114062 | 0.0188884 |
| %-Ω | -0.006% | -0.003% | 0.014% |
| Δ-ω [°] |  | 0.0116644 | -0.0189304 |
| %-ω | 0.010% | 0.008% | -0.035% |
| Δ-M [°] |  | -0.0912437 | -0.0000397 |
| %-M | 0.000% | -0.250% | 0.000% |
| Δ-MOID [AU] |  | 0.0000008 | -0.0000037 |
| %-MOID | -0.001% | 0.064% | -1.040% |

*Table 4. NEODyS-NEOPROP results comparison (2012DA14, 2007VK184, 2004DC)*

|  | 2012DA14 | 2007VK184 | 2004DC |
|---|---|---|---|
| *Epoch [JD]* | 2456275.0 | 2456400.5 | 2456400.5 |
| Δ-a [AU] | -0.0000003 | -0.0000100 | 0.0000000 |
| %-a | 0.000% | -0.001% | 0.000% |
| Δ-e [-] | -0.0000079 | -0.0000031 | 0.0000000 |
| %-e | -0.007% | -0.001% | 0.000% |
| Δ-i [°] | 0.0000266 | 0.0004886 | -0.0004472 |
| %-i | 0.000% | 0.040% | -0.002% |
| Δ-Ω [°] | 0.0050332 | -0.0784726 | 0.0047278 |
| %-Ω | 0.003% | -0.031% | 0.006% |
| Δ-ω [°] | -0.0042552 | 0.0785228 | -0.0049840 |
| %-ω | -0.002% | 0.107% | -0.003% |
| Δ-M [°] | 0.0003334 | 0.0076807 | -0.0000278 |
| %-M | 0.003% | 0.005% | 0.000% |
| Δ-MOID [AU] | -0.0000342 | 0.0000012 | -0.0000043 |

| %-MOID | 0.601%* | 0.184% | -0.047% |

*computed with the Numerical module

For the asteroid 2012DA14, the value reported as "MOID" is actually the close approach distance reached on the 15th of February [5]. The MOID value found on NEODyS was not used, since it was based on the new orbital elements determined after this Earth close approach (at epoch 2456400.5) and not on the reference elements used in this case (at epoch 2456275.0, before the close approach date). This is why the closest Earth approach distance was used as MOID.
To check this value (which is not a MOID) the Numerical module was used, instead of the Analytical one, to propagate the determined initial state until the day of the close approach. The computed numerical minimum close approach distance was equal to 0.00022899 AU, only 0.6% bigger than the reference value.

### 4.2. Numerical Module

The asteroid (99942) Apophis has been used to test the Numerical Module for a long-term propagation. Its orbital elements (previously determined) have been propagated until 14-04-2029 in order to cover its next Earth approach (on the 13-04-2029) and its consequent Moon close approach. Only these two close approaches will be evaluated, since NEOPROP records only close approaches below 0.005 AU or below 100 times the radius of each celestial body. The two mentioned close approaches are the first ones which match this requirement and they require about 25 years of numerical propagation, which is enough to evaluate the performance of the tool. Since the initial state used for the propagation comes from the Analytical module, this test will allow evaluating the overall performance of the NEOPROP software.
The JPL Small-Body Database was used as reference for the close approach distance [4], while the Horizons-JPL system was used as reference for the final state (as Cartesian elements) [3].
The final state and the close approaches distances (and epochs) are in line with the expected values. The results comparison is showed in the next table.

*Table 5. JPL Horizons-NEOPROP results comparison (99942)*

|  | Δ-Value | %-Value |
|---|---|---|
| x [km] | -2.139E+03 | 0.00% |
| y [km] | -4.366E+01 | 0.00% |
| z [km] | 5.573E+03 | -0.02% |
| Earth Distance [AU] | 1.366E-06 | 0.53% |
| Moon Distance [AU] | 8.743E-08 | 0.01% |

In the next table all the close approaches (from [4]) between 2006 and 2029 are reported, while in table Table 7 the data obtained by NEOPROP is summarized and compared to the reference values. The results of the close approaches are in line with the expectations.

*Table 6. Reference Apophis Close Approaches [4]*

| Time (TDB) | Time Uncertainty (hh:mm) | Body | Distance (AU) |
|---|---|---|---|
| 2006-Apr-10 23:49 | < 00:01 | Earth | 0.202819844 |
| 2013-Jan-09 11:42 | < 00:01 | Earth | 0.096661113 |
| 2016-Apr-24 02:49 | < 00:01 | Venus | 0.078241899 |
| 2020-Oct-12 08:37 | < 00:01 | Earth | 0.216276061 |
| 2021-Mar-06 01:14 | < 00:01 | Earth | 0.11265166 |
| 2024-Mar-07 15:45 | < 00:01 | Venus | 0.124432355 |
| 2029-Apr-13 21:46 | < 00:01 | Earth | 0.000256194 |
| 2029-Apr-14 14:33 | < 00:01 | Moon | 0.000636262 |

*Table 7. Apophis Close Approaches computed by NEOPROP*

| Time (TDB) | Body | Distance (AU) | Δ-Distance (AU) | %-Distance (AU) |
|---|---|---|---|---|
| 2006-Apr-10,23:49 | Earth | 0.202820223 | 3.794E-07 | 0.000% |
| 2013-Jan-09,11:44 | Earth | 0.096665441 | 4.328E-06 | 0.004% |
| 2016-Apr-24,02:50 | Venus | 0.078240242 | -1.656E-06 | -0.002% |
| 2020-Oct-12,08:36 | Earth | 0.21627714 | 1.078E-06 | 0.000% |
| 2021-Mar-06,01:16 | Earth | 0.112651436 | -2.238E-07 | 0.000% |
| 2024-Mar-07,15:45 | Venus | 0.124438646 | 6.291E-06 | 0.005% |
| 2029-Apr-13,21:46 | Earth | 0.000254828 | 1.366E-06 | 0.533% |
| 2029-Apr-14,14:33 | Moon | 0.000636175 | 8.743E-08 | 0.014% |

This test case was also used to validate the implemented integrators and to compare their performance. The last test case has been rerun with each integrator using the same set of inputs. Not all the integrator settings are used by each integrator (e.g. the fixed step-size integrators ignore the minimum step-size info). These settings are summarized here:

*Table 8. Integrators settings*

| Parameter | Value |
|---|---|
| Max step size | 8640.0 s |
| Min step size | 0.01 s |
| Local Tolerance | 1.0E-13 |
| Output interval | 86400.0 s |

The results have been compared with the output obtained in the previous test. Since the Gauss-Jackson 8 Adapted was used, this will be our reference.

In the next table the results are collected. Only the Earth and Moon MOIDs have been compared, since these represent the most sensitive data. In this way it was also possible to reduce a bit the output to be shown.

*Table 9. Integrators comparison*

| | Δ-MOID Earth | %-MOID Earth | Δ-MOID Moon | %-MOID Moon |
|---|---|---|---|---|
| **RK45** | 1.8129E-07 | 0.07% | 9.5067E-08 | 0.01% |
| **DP8** | 0.0000E+00 | 0.00% | 0.0000E+00 | 0.00% |
| **RK853** | -4.5716E-08 | -0.02% | 9.6784E-08 | 0.02% |
| **RK4** | -2.0973E-05 | **-8.26%** | -5.9186E-06 | -0.93% |
| **RK4_adapt** | 1.5362E-09 | 0.00% | 2.2645E-09 | 0.00% |
| **GJ8** | -2.4058E-05 | **-9.47%** | 3.6572E-05 | **5.73%** |
| **GJ8_adapt** | 0.0000E+00 | 0.00% | 0.0000E+00 | 0.00% |
| **GJ8_S_adapt** | 2.1324E-12 | 0.00% | -4.8585E-12 | 0.00% |

All the integrators have very similar results, except for the fixed step-size integrators: RK4 and GJ8. Their worse performance is intrinsic to the nature of these integrators. Since they lack of any control on the local error, they are not able to tackle situations where the perturbations become quite big (e.g. during a close approach with another celestial body). In order to improve their accuracy the fixed step-size should be reduced increasing the computational time. That was the reason why the "adapted" fixed step-size integrators have been developed within this project and seem to perform better than their original integration schemes.

A last test case was run for the 2007VK184 asteroid. The results of the orbit determination process have been already showed in paragraph 4.1. According to [6], 2007VK184 has a possible Earth impactor in 2048 (see details in the table below).

*Table 10. 2007VK184 Earth Impactor data [6]*

| Time (TDB) | σ (LOV) | Distance (Earth Radius) |
|---|---|---|
| 2048-Jun-03 02:08 | 1.29 | 0.92 |

Therefore, the Analytical Module has been rerun in order to generate VAs along the LOV. 2000 VAs have been generated and a maximum σ of 2 has been specified. Then the VAs has been passed to the Numerical module and propagated till 2048-Jun-04.

Also NEOPROP detected an Earth impactor at exactly the same epoch. The minimum Earth distance is slightly different from the reference value; maybe due to different values for the physical properties of the NEO (the ones used by NEODyS are unknown). Also the σ value of the VA is not the same; maybe due to a different definition of LOV adopted.

*Table 11. 2007VK184 Earth Impactor data NEOPROP*

| Time (TDB) | σ (LOV) | Distance (Earth Radius) |
|---|---|---|
| 2048-Jun-03 02:08 | 1.40 | 0.98 |

### 5. CONCLUSIONS

The NEOPROP tool is able to determine the orbital elements of a NEO starting from a set of observations and to propagate them in order to compute the minimum close approach distances to any planet. Particular focus has been put on determining impact risk with the Earth. The analytical MOID can be computed and used as main driver for a preliminary analysis. If an advanced analysis is requested, a set of VAs can be computed and propagated in order to find out if an Earth impact might occur in the future.

This software has been tested with several different NEOs and all the results are within 1% of their reference (NEODyS, JPL Small-Body Database and JPL-Horizons).

### 6. REFERENCES

1. Asteroid 2012 DA14 – Earth Flyby Reality Check. Online at http://www.nasa.gov/topics/solarsystem/features/asteroidflyby.html#TRAJECTORY%20DIAGRAMS (as of 13-03-2013)

2. NEODyS. Online at http://newton.dm.unipi.it/neodys/index.php?pc=0 (as of 13-03-2013)
3. JPL-HORIZONS. Online at http://ssd.jpl.nasa.gov/horizons.cgi#top (as of 22-01-2013)
4. Apophis Close Approaches. Online at http://ssd.jpl.nasa.gov/sbdb.cgi?sstr=apophis;orb=0;cov=0;log=0;cad=1#cad (as of 12-03-2013)
5. 2012DA14 Close Approaches. Online at http://ssd.jpl.nasa.gov/sbdb.cgi?sstr=2012%20DA14;orb=1;cov=0;log=0;cad=1#elem (as of 13-03-2013)
6. 2007VK184 Impactor Table. Online at http://newton.dm.unipi.it/neodys/index.php?pc=1.1.2&n=2007VK184 (as of 13-03-2013)